\documentclass[journal]{IEEEtran}
\usepackage{upgreek}

\ifCLASSINFOpdf
  \usepackage[pdftex]{graphicx}
  \graphicspath{{.}}
  \DeclareGraphicsExtensions{.pdf,.jpeg,.png}
\else
  \usepackage[dvips]{graphicx}
  \graphicspath{{../eps/}}
  \DeclareGraphicsExtensions{.eps}
\fi
\usepackage[caption=false,font=footnotesize]{subfig}
\newcommand{\unit}[2]{$#1\,\mathrm{#2}$}

\begin{document}
%
\title{Performance of a First-Level Muon Trigger with High Momentum Resolution Based on the ATLAS MDT Chambers for HL-LHC}
%
%
%

\author{P. Gadow, O.~Kortner, S. Kortner, H. Kroha, F. M\"uller, R. Richter\\ \textit{Max-Planck-Institut f\"ur Physik, Munich}}


%
%

\pagenumbering{gobble} 
%



\maketitle

\begin{abstract}
\boldmath
Highly selective first-level triggers are essential to exploit the full physics potential of the ATLAS experiment at High-Luminosity LHC (HL-LHC). The concept for a new muon trigger stage using the precision monitored drift tube (MDT) chambers to significantly improve the selectivity of the first-level muon trigger is presented. It is based on fast track reconstruction in all three layers of the existing MDT chambers, made possible by an extension of the first-level trigger latency to \unit{6}{\mu s} and a new MDT read-out electronics required for the higher overall trigger rates at the HL-LHC. 
Data from $pp$-collisions at $\sqrt{s} = 8\,\mathrm{TeV}$ is used to study the minimal muon transverse momentum resolution that can be obtained using the MDT precision chambers, and to estimate the resolution and efficiency of the MDT-based trigger. A resolution of better than $4.1\%$ is found in all sectors under study. With this resolution, a first-level trigger with a threshold of \unit{18}{GeV} becomes fully efficient for muons with a transverse momentum above \unit{24}{GeV} in the barrel, and above \unit{20}{GeV} in the end-cap region. 

\end{abstract}


%
\IEEEpeerreviewmaketitle

\section{Introduction}
\begin{figure}[tbp]
\centering
\includegraphics[width=0.7\columnwidth]{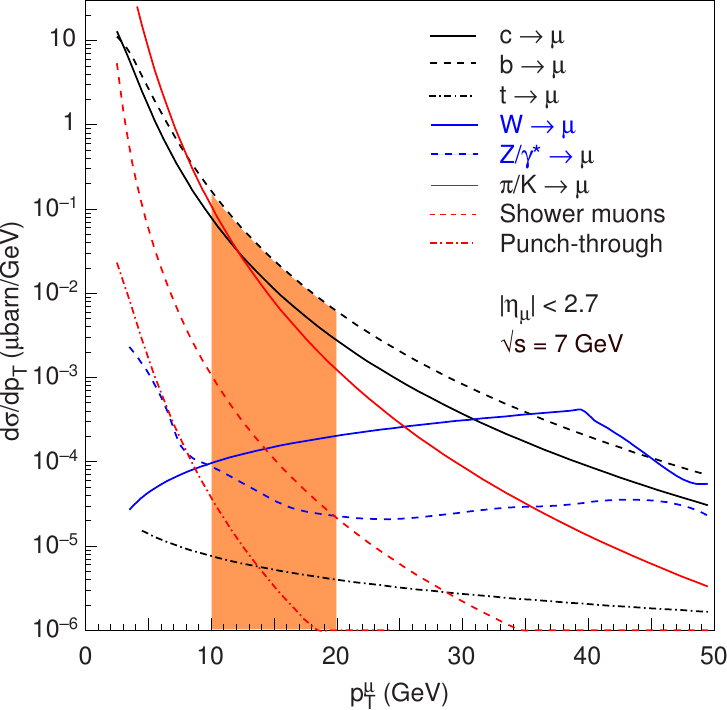}
\caption{Differential cross-section $d\sigma/dp_\mathrm{T}$ for the dominant muon production processes as function of the muon transverse momentum at a centre-of-mass energy of $\sqrt{s} = 7\,\mathrm{TeV}$ in the pseudo-rapidity range $|\eta_\upmu| < 2.7$~\cite{cite:mtdr}. The orange area marks the momentum range which is critical to be suppressed by a Level-1 MDT trigger.}
\label{fig:1_muon_xsec}
\end{figure}
Highly selective first-level triggers are essential to exploit the full physics potential of the ATLAS experiment at the HL-LHC where the instantaneous luminosity will exceed the LHC Run 1 instantaneous luminosity by almost an order of magnitude. In order to cover the interesting electroweak physics processes, the goal for the Level-1 muon trigger is to maintain a trigger for single muons that is fully efficient at \unit{20}{GeV} without the need for prescaling. Figure~\ref{fig:1_muon_xsec} shows the differential cross-section $d\sigma/dp_\mathrm{T}$ for the dominant muon production processes as function of the muon transverse momentum at a centre-of-mass energy of $\sqrt{s} = 7\,\mathrm{TeV}$ in the pseudo-rapidity range $|\eta_\upmu| < 2.7$. It is obvious that in order to suppress the steeply rising cross-section for $p_\mathrm{T} < 20\,\mathrm{GeV}$, a very good momentum resolution for the muon trigger and a sharp turn-on of the trigger efficiency is required. 

\begin{figure}[tbp]
\centering
\includegraphics[width=0.9\columnwidth]{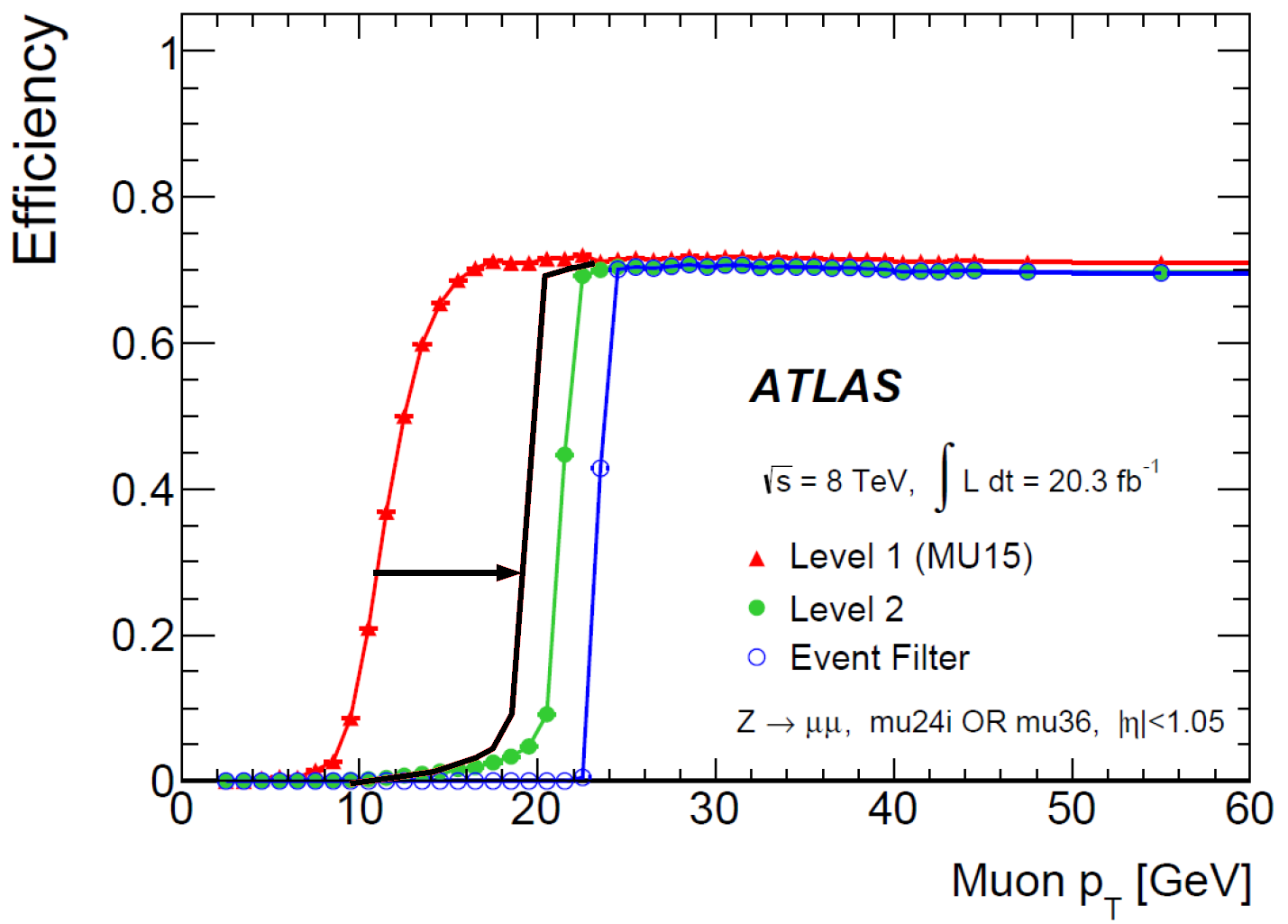}
\caption{Trigger efficiency as function of the muon transverse momentum~$p_\mathrm{t}$  at the three trigger stages Level-1, Level-2 and Event Filter~\cite{cite:tdaq}. The black line has been added as illustration for the envisaged Level-1 MDT trigger efficiency.}
\label{fig:1_efficiency}
\end{figure}
The ATLAS experiment plans to increase the acceptable trigger rate of the first two trigger levels to \unit{1}{MHz} at \unit{6}{µs} latency and \unit{400}{kHz} at \unit{30}{µs} latency, respectively. This requires to replace the electronics for the muon trigger decision, as well as the read-out electronics of the muon trigger chambers and the precision monitored drift tube (MDT) chambers. The replacement of the precision chamber read-out electronics will make it possible to include their data in the first-level trigger decision and to increase the selectivity of the first-level muon trigger~\cite{cite:sMDT_upgrade,cite:sMDT_HLLHC}. The MDT-based muon trigger will be implemented at Level-0 of the new ATLAS trigger system for HL-LHC, which will replace the current Level-1. 

The improvements on the current ATLAS Level-1 muon trigger efficiency is illustrated in Figure~\ref{fig:1_efficiency}, which shows the efficiency at the three trigger stages Level-1, Level-2 and Event Filter. The resolution of the Level-1 MDT trigger is expected to be very close to the one of the Level-2, resulting in a turn-on as indicated in the figure and a substantial rate reduction from the existing Level-1 trigger~\cite{cite:scoping}.

Two possible concepts for a Level-1 MDT trigger are currently under study.
In the two-station concept, the two outer of the three layers of the MDT chambers are used. The muon momentum is determined from the deflection angle in the magnetic field comparing the two measured track segments in the layers. The rate reduction for the two-station concept has been determined using ATLAS data at $\sqrt{s} = 8\,\mathrm{TeV}$. It is found to be in the order of $50\%$ across the entire $\eta$-coverage of the muon spectrometer~\cite{cite:scoping}. 
The three-station concept makes use of all three layers of the MDT precision chambers and relies on the excellent spatial resolution of the MDT chambers, which exceeds the angular resolution by a factor of about two. Hence, the performance of the three-station concept is expected to be comparable to the one from the Level-2 trigger. In this contribution, the resolution and efficiency of the three-station concept is studied, using data from the ATLAS data taking at $\sqrt{s} = 8\,\mathrm{TeV}$. 


\section{The ATLAS muon spectrometer}
\label{sec:atlas}
The ATLAS detector consists of a tracking system in a \unit{2}{T} axial magnetic field up to a pseudorapidity\footnote{ATLAS uses a right-handed coordinate system with its origin at the nominal interaction point (IP) in the centre of the detector and the $z$-axis along the beam pipe. The $x$-axis points from the IP to the centre of the LHC ring, and the $y$-axis points upward. Cylindrical coordinates $(r,\phi)$ are used in the transverse plane, $\phi$ being the azimuthal angle around the $z$-axis. The pseudorapidity is defined in terms of the polar angle $\theta$ as $\eta=-\ln\tan(\theta/2)$.} of $|\eta| < 2.5$, sampling electromagnetic and hadronic
calorimeters up to $|\eta| < 4.9$, and the muon spectrometer in a \unit{0.5}{T} azimuthal magnetic field provided by a system of air-core toroidal magnets. A detailed description of the ATLAS detector can be
found elsewhere~\cite{cite:atlas}.

The muon spectrometer consists of the barrel region with $|\eta| < 1.05$, and the end-cap region with $1.4 < |\eta| < 2.7$. The magnet system is divided accordingly, where both barrel and end-cap toroid magnets consist of eight coils each, resulting in an eightfold symmetry. In the barrel, three layers of MDT chambers are installed. In the area between the magnet coils, the so-called \textit{large sectors}, the distance to the beam axis are \unit{4.9}{m}, \unit{7.1}{m}, and \unit{9.5}{m}, respectively. Within the coils, in the so-called \textit{small sectors}, the chambers are placed in a distance of \unit{4.6}{m}, \unit{8.1}{m}, and \unit{10.6}{m}, respectively. While the magnetic field strength is higher for the small sectors, muons in this region suffer from multiple interactions with the additional passive material of the magnets and their support structure. The middle and outer layer in the barrel are equipped with additional resistive plate chambers (RPCs) dedicated to fast triggering for the current Level-1 trigger.
The end-cap consists of three discs with large sectors at a distance of $\pm$\unit{7.7}{m}, $\pm$\unit{14.3}{m}, and $\pm$\unit{21.4}{m} and small sectors at a distance of $\pm$\unit{7.3}{m}, $\pm$\unit{13.9}{m}, and $\pm$\unit{21.8}{m} from the interaction point. The New Small Wheel (NSW) replaces the innermost layer of the current end-cap of the muon detector for the HL-LHC and will use Micromegas. The middle and outermost layer of the end-cap are build from MDT chambers. Three layers of dedicated thin gap chambers (TGCs) for fast triggering are installed around the middle layer of the end-cap.

Due to the large size and weight of the detector, some areas in the $\eta$-$\phi$ space are not fully covered by muon detectors, in particular the transition between the barrel and the end-cap at $1.05 < |\eta| < 1.4$, as well as the support structure of the detector at $3.53 < \phi < 5.11$. Hence, these regions have been excluded from the study. 

\section{Trigger concept}
\begin{figure}[tbp]
\centering
\includegraphics[width=0.9\columnwidth]{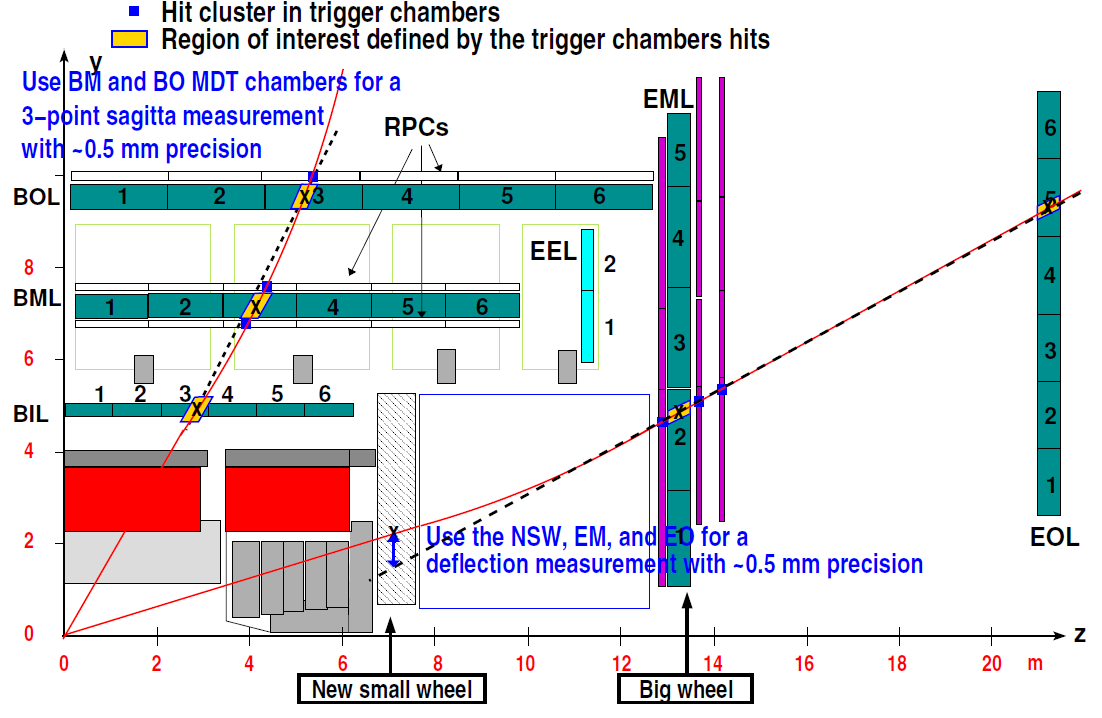}
\caption{Side view of one quadrant of the ATLAS detector for illustration of the 3-station concept. Muon tracks are indicated by red lines. The hit clusters and regions of interest in the trigger chambers are marked by blue and yellow markers, respectively. }
\label{fig:2_concept}
\end{figure}
The trigger concept is illustrated in Figure~\ref{fig:2_concept}, which shows a side view of one quadrant of the ATLAS detector. The Level-1 MDT trigger relies on the Regions of Interest (RoI) defined by the dedicated existing Level-1 trigger chambers. The RoIs are built from hits in the RPCs and TGCs, which can be extrapolated to all three layers of the precision MDT chambers using their position and momentum information. 

In the barrel, the muon track is measured at the three space points around the RoIs. The transverse momentum of the muon is determined from the curvature of the muon track in the middle layer by calculating the sagitta $\Delta s$ from the projection of the track on the sector normal in $\phi$. The sagitta is defined by the shortest distance of the measured track segment in the middle layer with respect to a straight line between the track segments in the inner and outer layer.

In the end-cap region, the muon track is measured at three space points given by NSW, middle end-cap (EM) and outer end-cap (EO). Due to the position of the end-cap magnet, the magnetic field is limited to the space between the inner and middle end-cap layer. Hence, the most sensitive method to determine the transverse momentum of the muon is from the curvature inside those two layers. The measured quantity is the distance $\Delta L$ between the track segment in the inner layer, and the extrapolated track from the measured track segments in the middle and outer layer of the end-cap.  

\section{Data analysis}
Data from the ATLAS data taking at $\sqrt{s} = 8\,\mathrm{TeV}$, corresponding to an integrated luminosity of $5.5\,\mathrm{fb}^{-1}$, is used for this study. 
The events are preselected by having at least one positive trigger decision among all the muon trigger items. No requirement on a particular Level-1 single muon trigger is imposed to minimise the statistical uncertainty on the data.

Two different muon reconstruction algorithms are employed in this study. The \textit{combined} algorithm makes use of the combined information from the inner tracking detectors and the muon spectrometer. It has the best transverse momentum resolution and hence is considered as reference algorithm. The \textit{standalone} algorithm reconstructs the muon transverse momentum with information from the muon spectrometer only. It represents the benchmark for the trigger algorithm. Quantities measured with the combined and standalone algorithm are indicated by the superscripts $^{CB}$ and $^{SA}$, respectively. 
Due to energy loss in the calorimeters, the standalone momentum is systematically lower than the combined momentum, therefore the standalone momentum has to be corrected by the most probable energy loss $\Delta E$ during the passage from the interaction point to the muon spectrometer.
The most probable energy loss of muons with $10\,\mathrm{GeV}<p_{\mathrm{T}}^{CB} < 30\,\mathrm{GeV}$ in large barrel sectors is $\Delta E_{\mathrm{B}} = 2.51\pm 0.001\,\mathrm{GeV}$, while the most probable energy loss in the end-cap sectors is $\Delta E_{\mathrm{EC}} = 0.936\pm0.001\,\mathrm{GeV}$.

The Level-1 MDT trigger algorithm is studied on basis of muon track segments as reconstructed in the individual layers of the MDT chambers. The track segments are required to have at least six hits, and each layer is required to have one track segment in the region of interest. In the overlap region between small and large sectors, information from only one sector type is used. For simplicity, only positively charged muons with $\eta > 0$ are considered. It has been verified that negatively charged muons and muons in the region $\eta < 0$ give similar results. Muon tracks with $p_{\mathrm{T}}^{SA} > 100\,\mathrm{GeV}$ are not considered in the analysis.
The transition region between barrel and end-cap as well as the region of the support structure has been excluded, as described in Section~\ref{sec:atlas}. For the end-cap, this study focuses on the region in $2.0 < \eta < 2.4$ as a guide for further investigation.
Since the NSW has not been installed yet in the selected data period, the information from the inner end-cap of MDT chambers is used.
  
\section{Resolution of the standalone algorithm}
\begin{figure}[tbp]
\centering
\includegraphics[width=0.9\columnwidth]{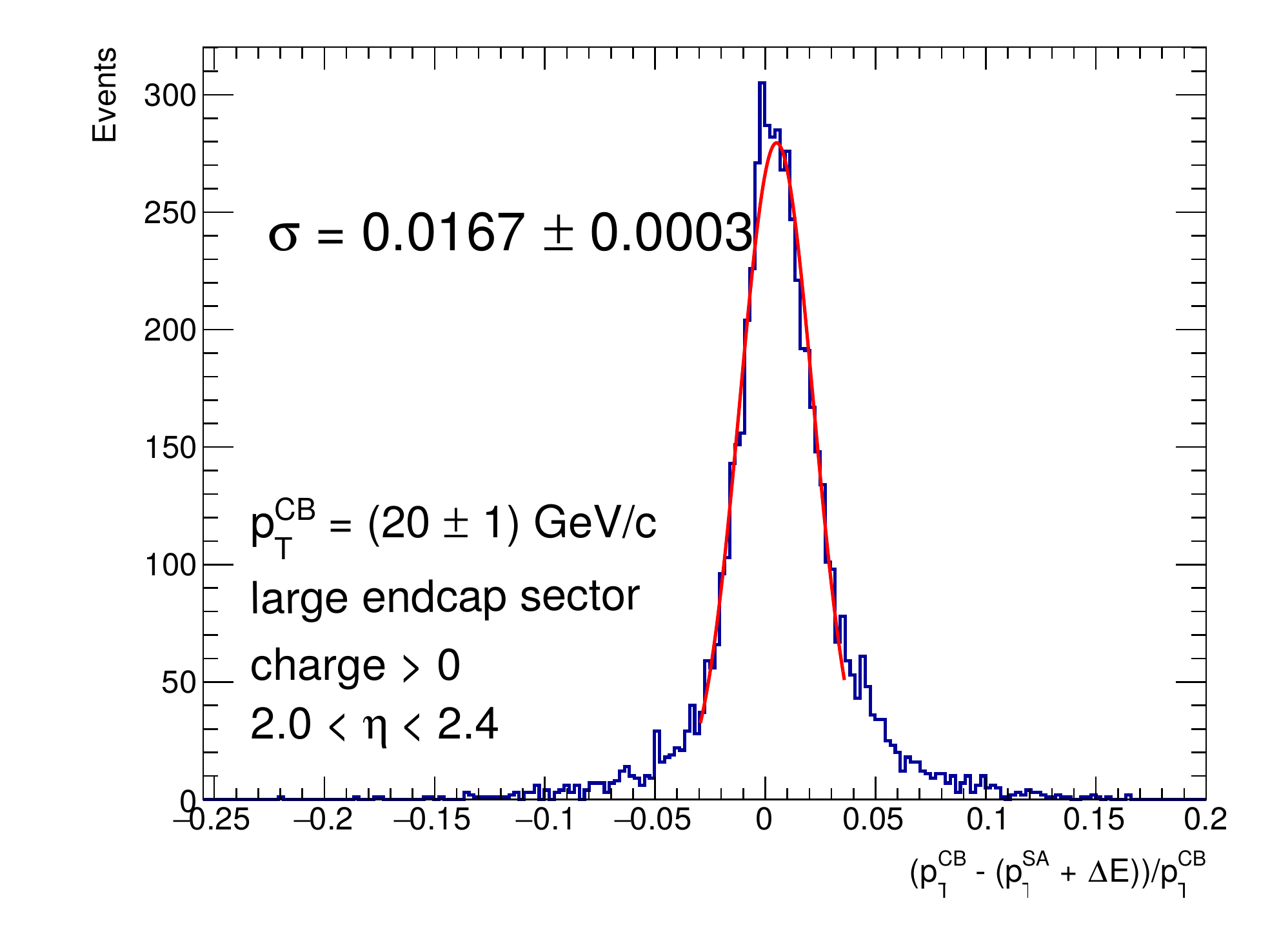}
\caption{Momentum resolution for positively charged muons in large sectors with $19\,\mathrm{GeV} < p_\mathrm{T}^\mathrm{CB} < 21\,\mathrm{GeV}$ and $2.0 < \eta < 2.4$.}
\label{fig:3_resolution}
\end{figure}

\begin{figure}[tbp]
\centering
\includegraphics[width=0.9\columnwidth]{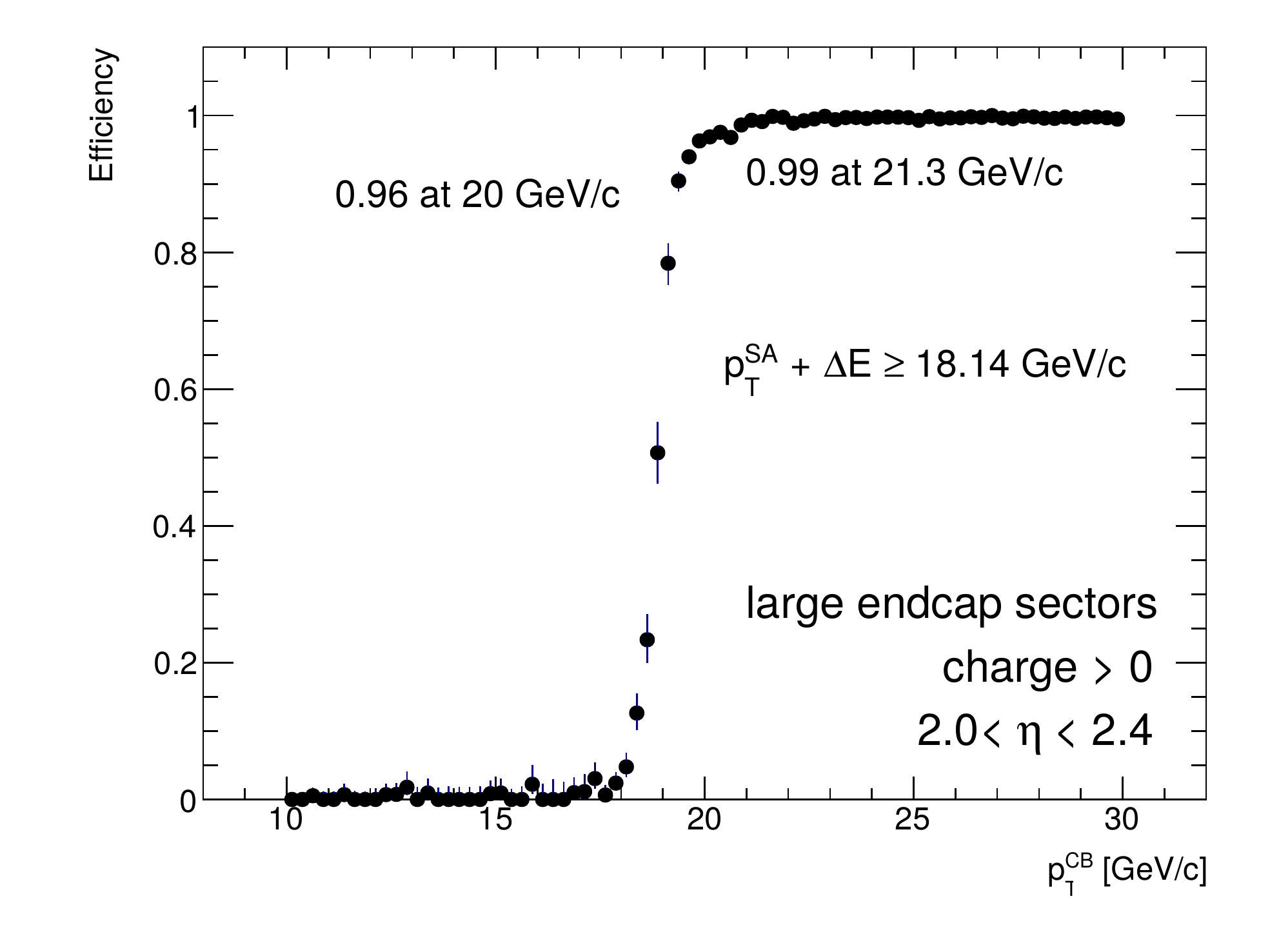}
\caption{Trigger efficiency for positively charged muons in large sectors at $2.0 < \eta < 2.4$ as function of the muon transverse momentum $p_\mathrm{T}^\mathrm{CB}$ reconstructed using the combined algorithm. A threshold of $p_\mathrm{T}^\mathrm{SA} = 18.14\,\mathrm{GeV}$ has been applied.}
\label{fig:3_efficiency_endcap_3sig}
\end{figure}
\begin{figure}[tbp]
\centering
\includegraphics[width=0.9\columnwidth]{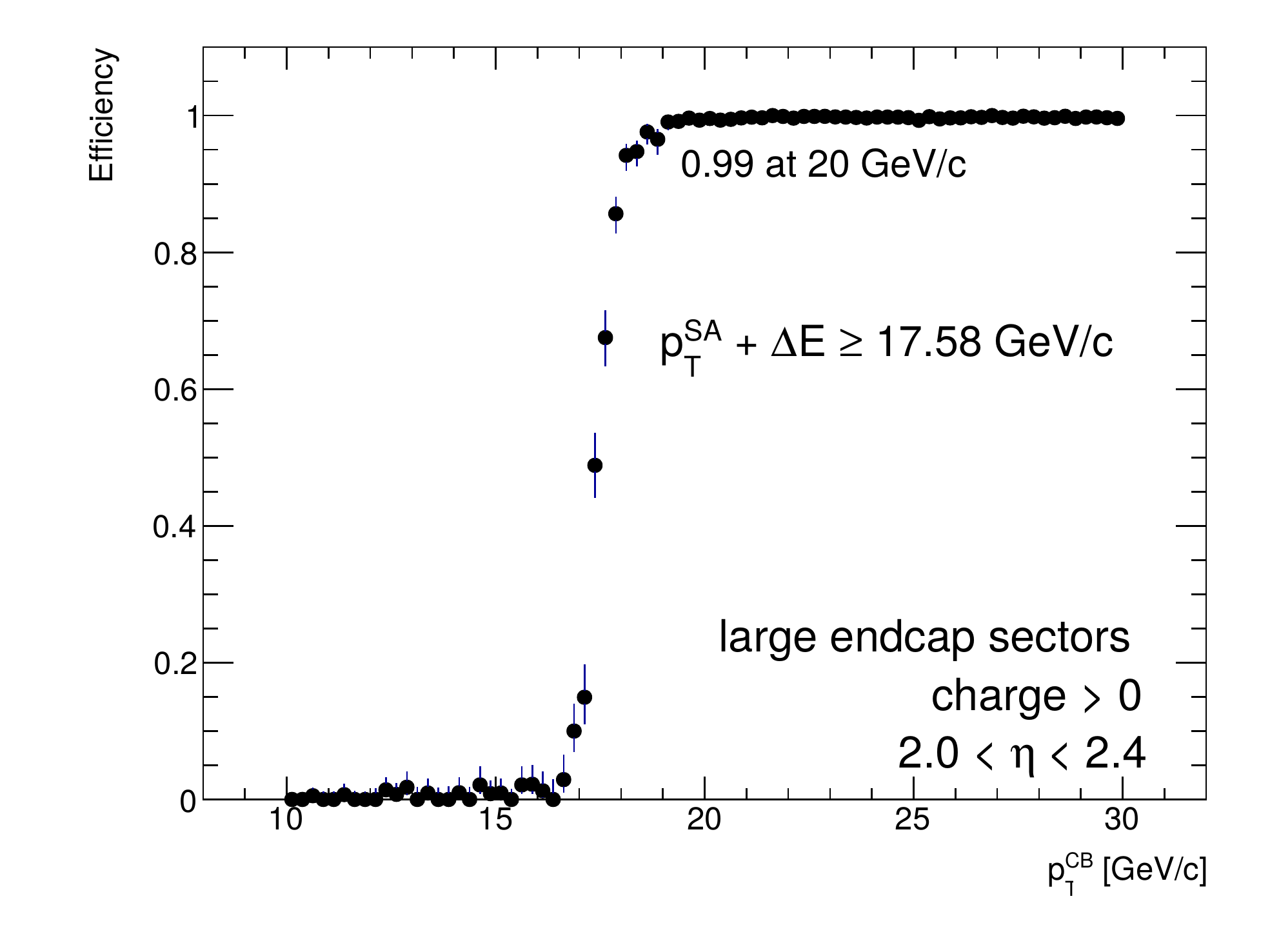}
\caption{Trigger efficiency for positively charged muons in large sectors at $2.0 < \eta < 2.4$ as function of the muon transverse momentum $p_\mathrm{T}^\mathrm{CB}$ reconstructed using the combined algorithm. A threshold of $p_\mathrm{T}^\mathrm{SA} = 17.58\,\mathrm{GeV}$ has been applied.}
\label{fig:3_efficiency_endcap_99}
\end{figure}

\begin{figure*}[tbp]
  \subfloat[\label{fig:4_calibration_pt_prefit}]{%
    \includegraphics[width=0.32\textwidth]{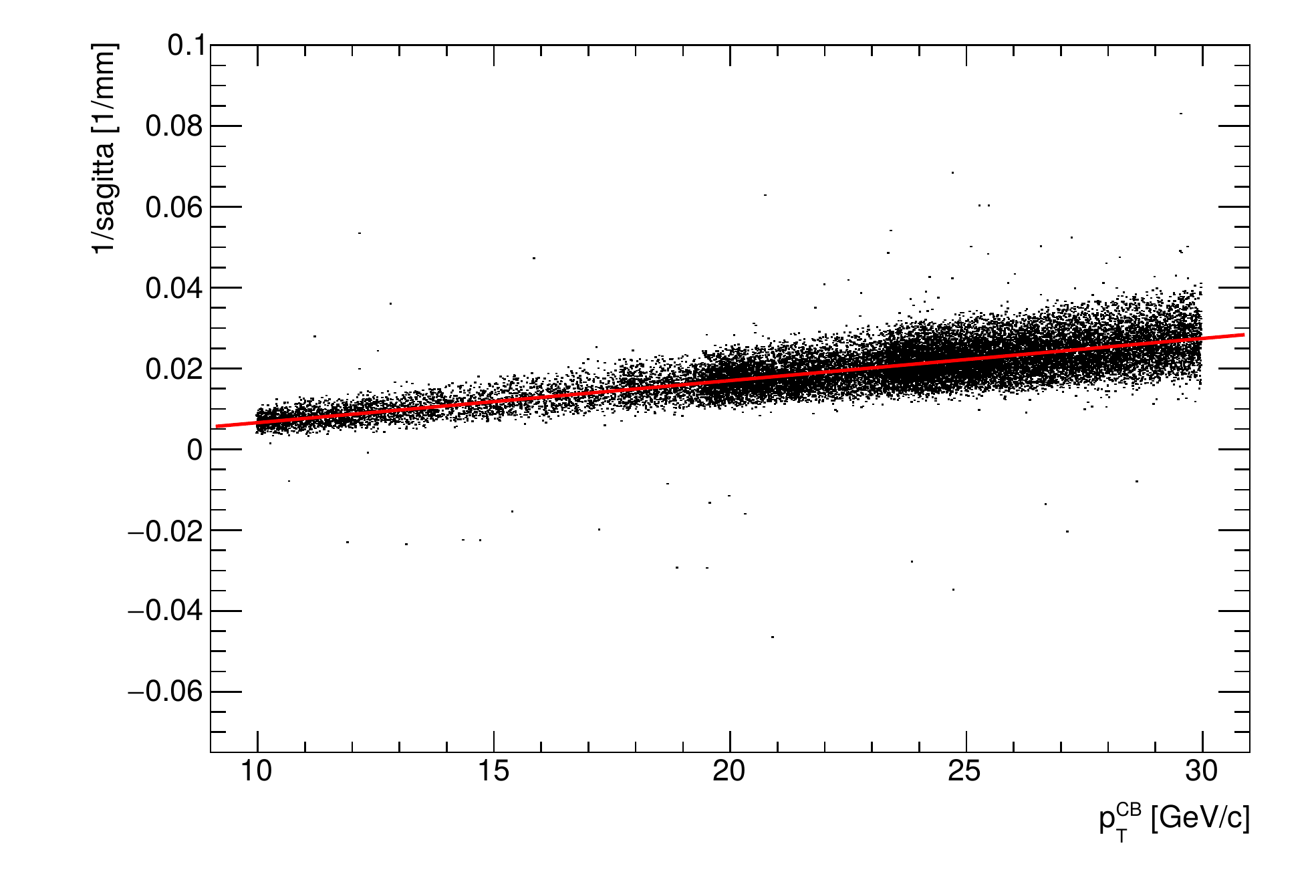}
  }
  \hfill
  \subfloat[\label{fig:4_calibration_pt_postfit}]{%
    \includegraphics[width=0.32\textwidth]{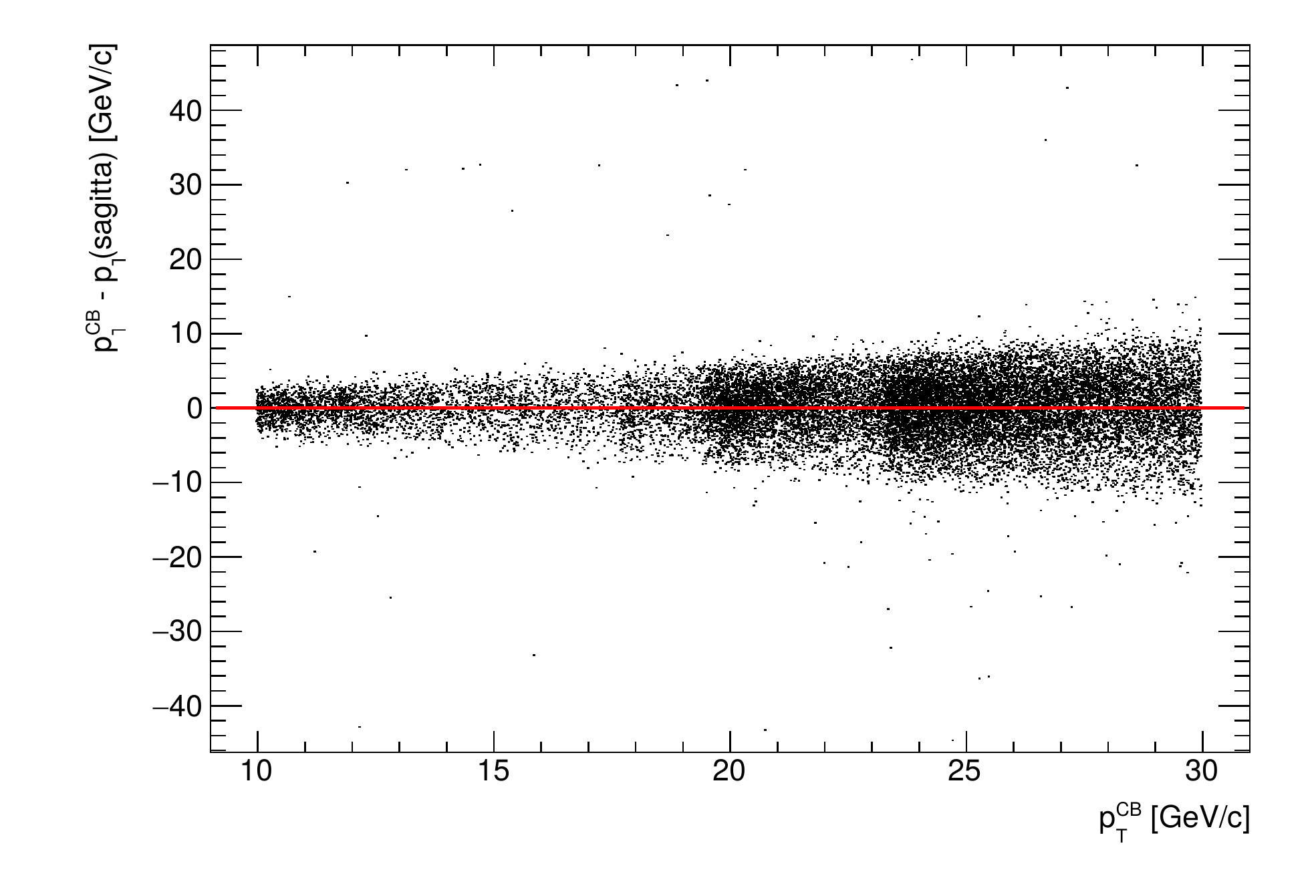}
  }
  \hfill
  \subfloat[\label{fig:4_calibration_phi_prefit}]{%
    \includegraphics[width=0.32\textwidth]{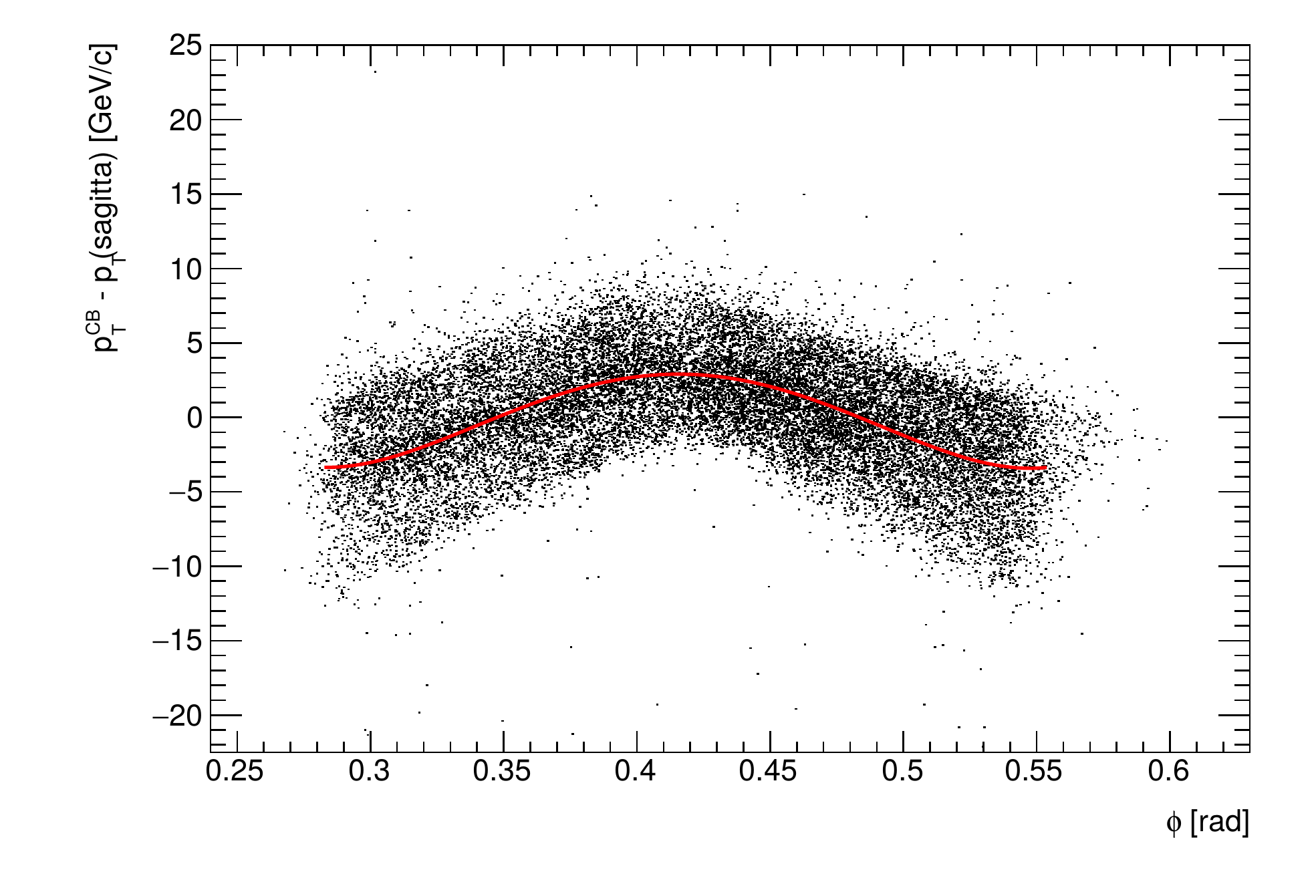}
  } \\
  \subfloat[\label{fig:4_calibration_phi_postfit}]{%
    \includegraphics[width=0.32\textwidth]{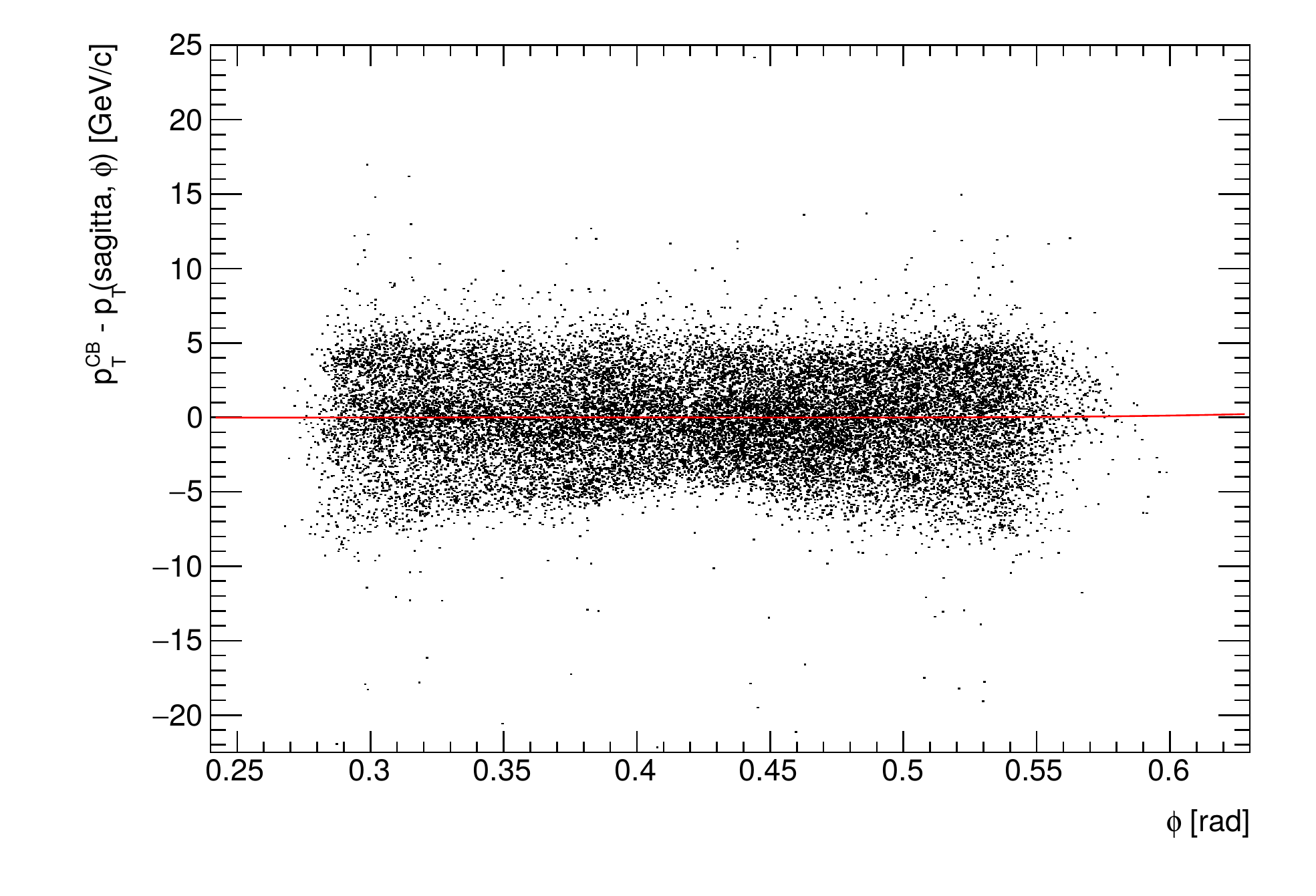}
  } 
  \hfill
  \subfloat[\label{fig:4_calibration_eta_prefit}]{%
    \includegraphics[width=0.32\textwidth]{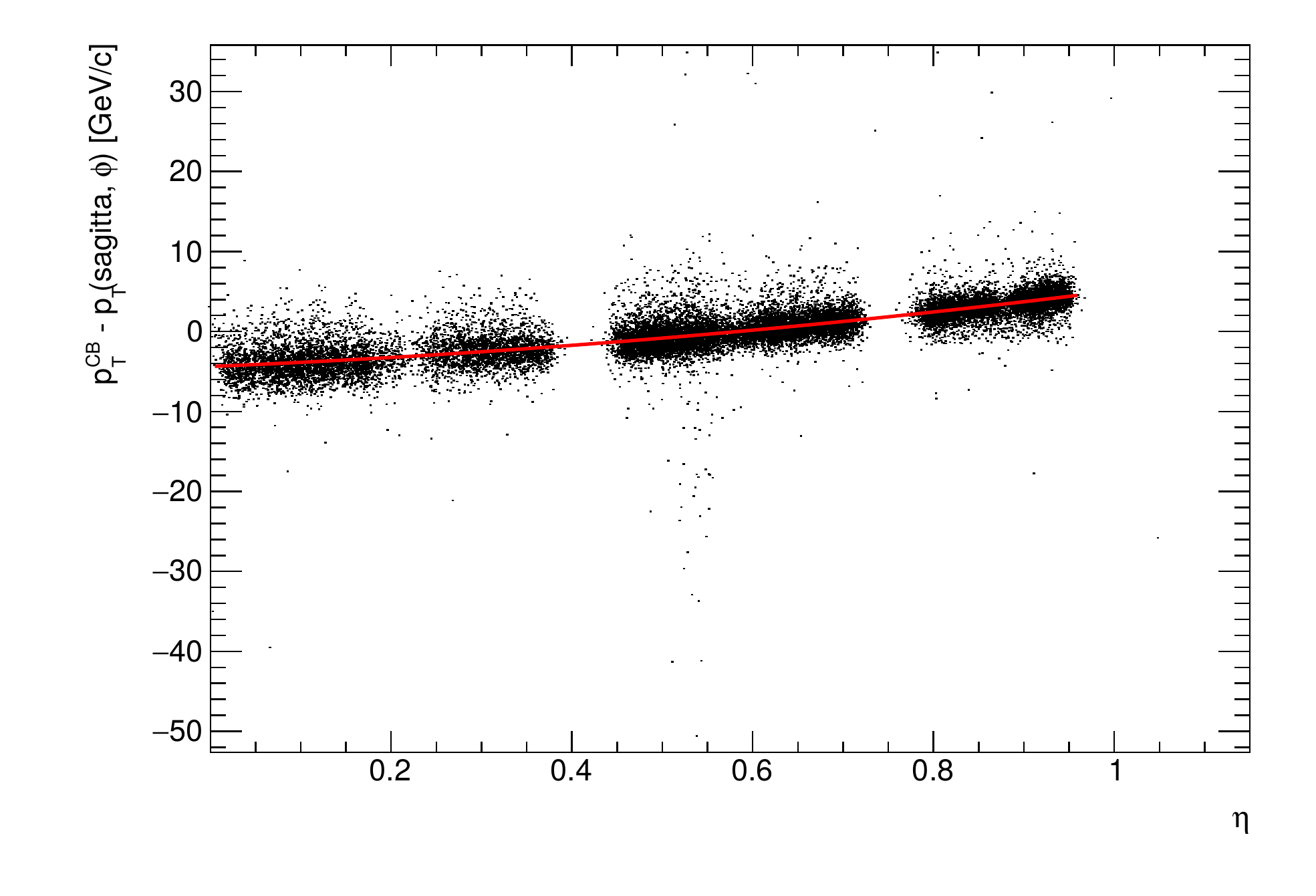}
  }
  \hfill
  \subfloat[\label{fig:4_calibration_eta_postfit}]{%
    \includegraphics[width=0.32\textwidth]{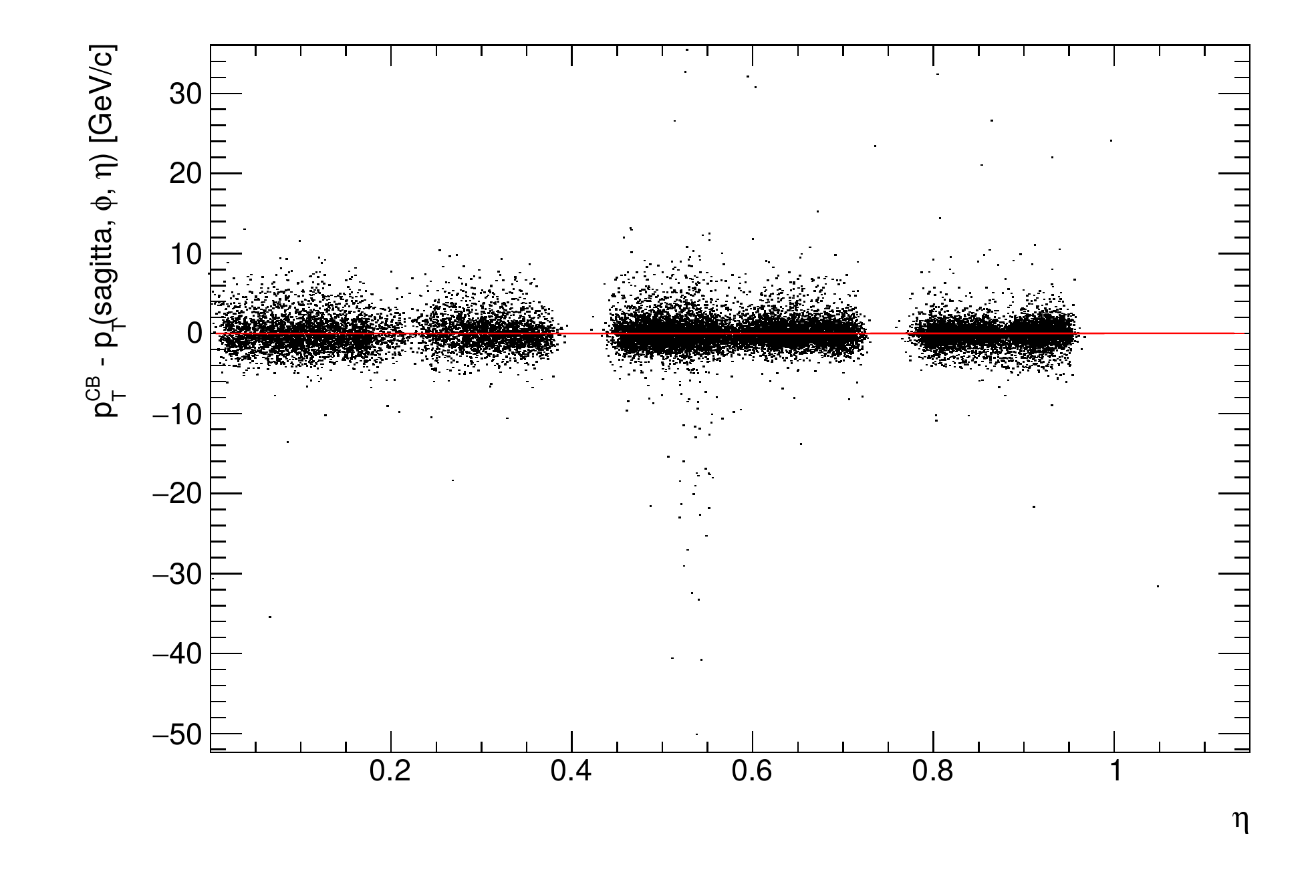}
  }
  \caption{Iterative method to determine the calibration function $p_\mathrm{T}(s,\phi,\eta)$. Data from positively charged muons in a small sector of the barrel region is used.
  }
  \label{fig:4_calibration}
\end{figure*}

The standalone algorithm makes use of only the muon spectrometer data, hence it can be used to give an estimate for an upper limit of the MDT trigger algorithm's performance. 
The trigger's efficiency is defined as the ratio of events passing the threshold over all events in a certain momentum range.
A trigger can be considered fully efficient once it reaches a plateau of
 $>99\%$. For a momentum resolution with a Gaussian shape, the $99\%$ point is reached at $3\sigma$ above the trigger threshold. 
In order for the trigger to be efficient at the efficiency point $p_\mathrm{T}^\mathrm{eff} = 20\,\mathrm{GeV}$, the corresponding threshold must be set at

\begin{equation}
\label{eq:3_pthresh}
p_\mathrm{T}^\mathrm{thresh} = p_\mathrm{T}^\mathrm{eff} - 3\sigma \cdot p_\mathrm{T}^\mathrm{eff}.
\end{equation}

A better momentum resolution translates to a more selective threshold.
For the barrel region, the expected resolution of the standalone algorithm for the relevant transverse momentum range around $p_\mathrm{T}^\mathrm{eff} = 20\,\textrm{GeV}$, the resolution is about $4\%$~\cite{cite:atlas}.
The momentum resolution for large end-cap sectors at \unit{20}{GeV} in the end-cap region $2.0 < \eta < 2.4$ is shown in Figure~\ref{fig:3_resolution}. Albeit the Gaussian core of the distribution is narrow, with a width of a standard deviation $\sigma = 0.0167 \pm 0.0003$, there are non-Gaussian tails, particularly in the barrel region, resulting from energy loss fluctuations.
As the energy loss fluctuations set a fundamental limit on the momentum resolution and hence on the convergence of the trigger efficiency at the plateau, being fully efficient at $p_\mathrm{T}^\mathrm{eff}$ comes at the cost of a reduced rate reduction.

One should note that in the end-cap region the combined reconstruction algorithm relies predominantly on the muon spectrometer data, resulting in a non-negligible correlation between the two values and a possible overestimation of the resolution.

The trigger efficiency for the large sectors in the end-cap region $2.0 < \eta < 2.4$ is shown in Figure~\ref{fig:3_efficiency_endcap_3sig}. 
The turn-on curve reaches $0.96$ efficiency at \unit{20}{GeV} and becomes fully efficient at \unit{21.3}{GeV}.
The fact that the trigger is not fully efficient at \unit{20}{GeV} using the $3\sigma$ estimation is caused by energy loss fluctuations, which are responsible for the non-Gaussian tails of the momentum resolution distribution.
In order to be fully efficient at \unit{20}{GeV} a lower threshold of $p_\mathrm{T}^\mathrm{thresh} =17.58\,\mathrm{GeV}$ has to be chosen, as shown in Figure~\ref{fig:3_efficiency_endcap_99}. Here, the turn-on curve is less steep compared to Figure~\ref{fig:3_efficiency_endcap_3sig}, resulting in a smaller rate reduction.

\section{Calibration procedure}

The sagitta $\Delta s$ as a measure for the path's curvature is inversely proportional to the particle's transverse momentum $p_\mathrm{T}$, which is expressed by the well-known relation
\begin{equation}
\Delta s = \frac{eBL^2}{8 p_\mathrm{T}}.
\label{eq_1}
\end{equation}
As the magnetic field $B(\phi, \eta)$ is not constant over the whole detector, a parametrization of $p_\mathrm{T}(\Delta s, \phi, \eta)$ is constructed using an iterated fit procedure by the means of Equation~\ref{eq_1}.
To account for local inhomogeneities, the barrel region is divided into eight large and eight small sectors evaluated independently.
In Figure~\ref{fig:4_calibration_pt_prefit} the inverse sagitta is shown against the combined momentum of the muon, which is used as a calibration measure for $p_\mathrm{T}$. A linear function $f(p_{\mathrm{T}}) = a_0 + a_1 \cdot p_\mathrm{T}$ is fitted to the data with the method of $\chi^2$ minimization to obtain 
\begin{equation}
p_\mathrm{T}(\Delta s) = \frac{1/\Delta s - a_0}{a_1}.
\end{equation}
After obtaining the fit parameters, the residuals show no strong $p_\mathrm{T}$ dependence, as can be seen in Figure~\ref{fig:4_calibration_pt_postfit}.
In a second iteration, shown in Figure~\ref{fig:4_calibration_phi_prefit}, the residual $p_\mathrm{T}^{CB} - p_\mathrm{T}(\Delta s)$ is shown against $\phi$ and a quartic polynomial $P_4(\phi) = \sum_{n=1}^4 p_n \phi^n$ is fitted to the data to obtain
\begin{equation}
p_\mathrm{T}(\Delta s, \phi) = \frac{1/\Delta s - a_0}{a_1} + P_4(\phi).
\end{equation}
The residuals are shown in Figure~\ref{fig:4_calibration_phi_postfit}, where  no strong dependence on $\phi$ is observed.
The remaining dependence on $\eta$ is shown in Figure~\ref{fig:4_calibration_eta_prefit}, where the residuals $p_\mathrm{T}^{CB} - p_\mathrm{T}(\Delta s, \phi)$ are shown against $\eta$. 
A quadratic polynomial $E_2(\eta) = \sum_{n=1}^2 p_n \eta^n$ is fitted to the data to obtain
\begin{equation}
p_\mathrm{T}(\Delta s, \phi, \eta) = \frac{1/\Delta s - a_0}{a_1} + P_4(\phi) + E_2(\eta).
\end{equation}
The residuals shown in Figure~\ref{fig:4_calibration_eta_postfit} prove that the parametrisation of $p_\mathrm{T}$ in terms of sagitta, $\phi$ and $\eta$ is sound.
The same methodology is applied in the end-cap using $\Delta L$ instead of $\Delta s$.


\section{Trigger performance}

\begin{figure}[tbp]
\centering
\includegraphics[width=0.9\columnwidth]{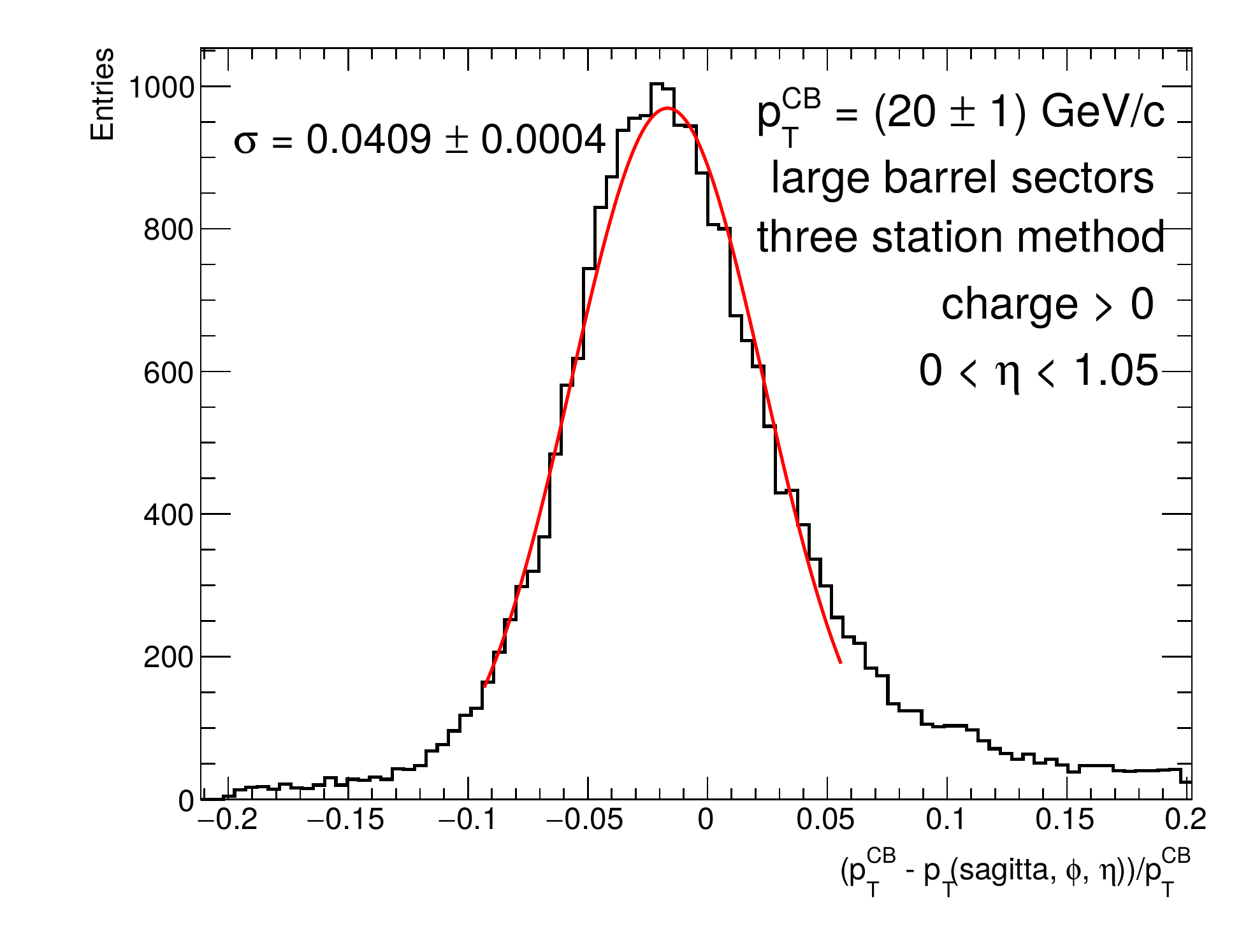}
\caption{Momentum resolution using the parametrisation $p_{\mathrm{T}}(\Delta s,\phi,\eta)$ for positively charged muons in large sectors with $19\,\mathrm{GeV} < p_\mathrm{T}^\mathrm{CB} < 21\,\mathrm{GeV}$ and $0.0 < \eta < 1.05$.}
\label{fig:5_resolution_barrel}
\end{figure}
\begin{figure}[tbp]
\centering
\includegraphics[width=0.9\columnwidth]{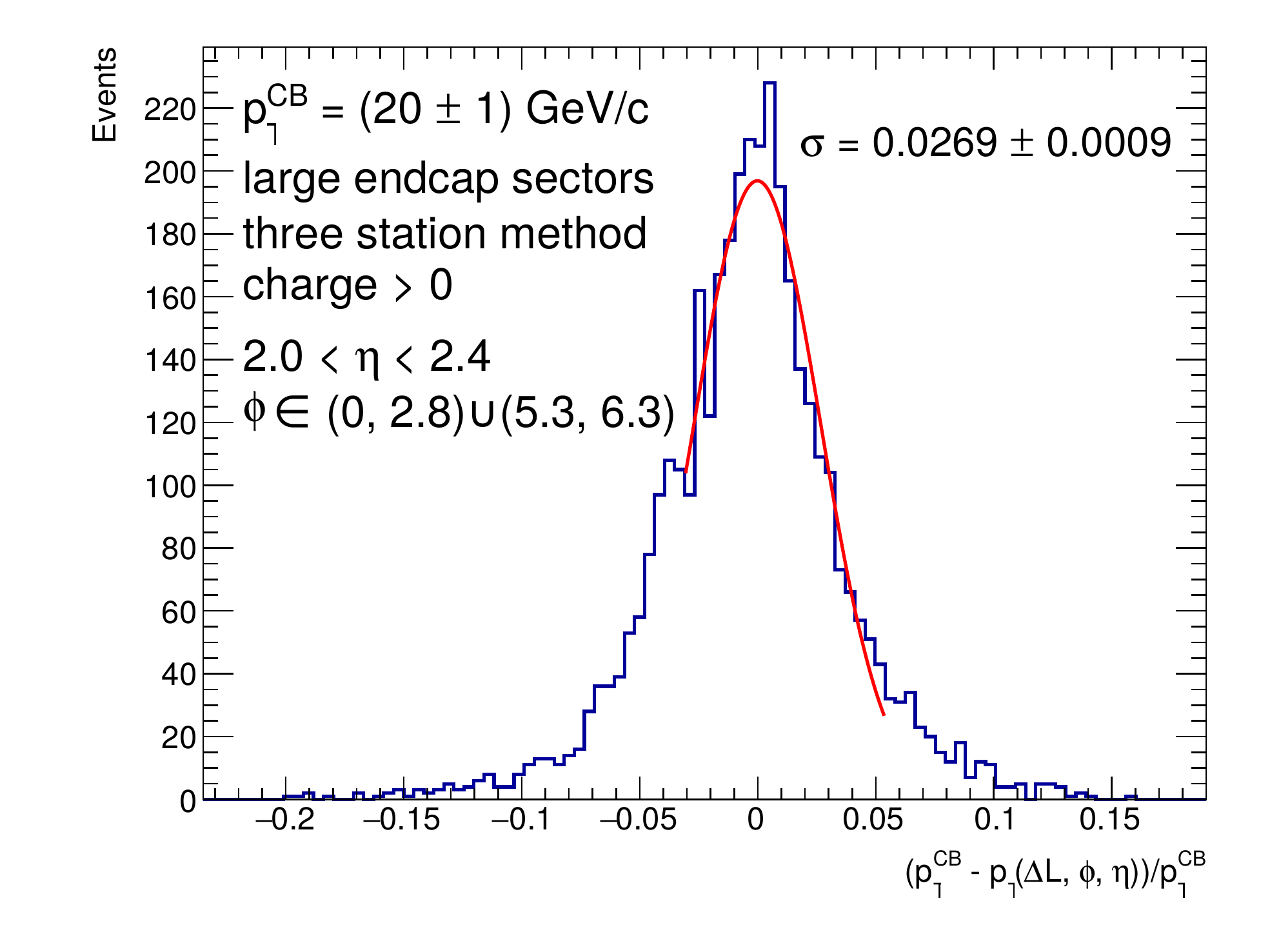}
\caption{Momentum resolution using the parametrisation $p_{\mathrm{T}}(\Delta L,\phi,\eta)$ for positively charged muons in large sectors with $19\,\mathrm{GeV} < p_\mathrm{T}^\mathrm{CB} < 21\,\mathrm{GeV}$, $\phi \in (0,2.8)\cup(5.28,6.28)$ and $2.0 < \eta < 2.4$.}
\label{fig:5_resolution_endcap}
\end{figure}
\begin{figure}[tbp]
\centering
\includegraphics[width=0.9\columnwidth]{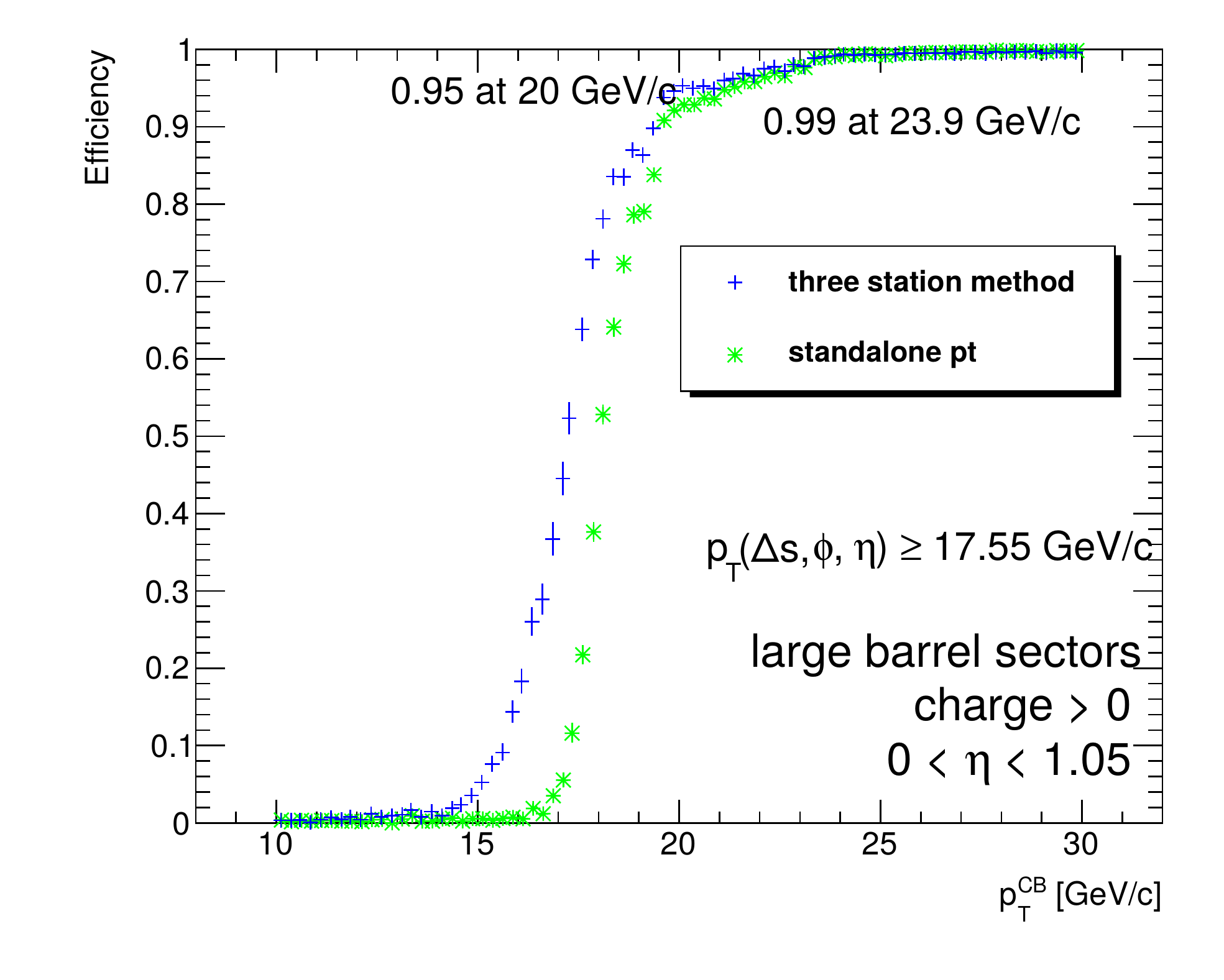}
\caption{Trigger efficiency for positively charged muons in large sectors at $0.0 < \eta < 1.05$ as function of the muon transverse momentum $p_\mathrm{T}^\mathrm{CB}$ reconstructed using the combined algorithm. A threshold of $p_\mathrm{T}^\mathrm{SA} = 17.55\,\mathrm{GeV}$ has been applied using the muon transverse momentum given by the parametrisation $p_{\mathrm{T}}(\Delta s,\phi,\eta)$ and by the energy-loss corrected standalone algorithm.}
\label{fig:5_efficiency_barrel}
\end{figure}
\begin{figure}[tbp]
\centering
\includegraphics[width=0.9\columnwidth]{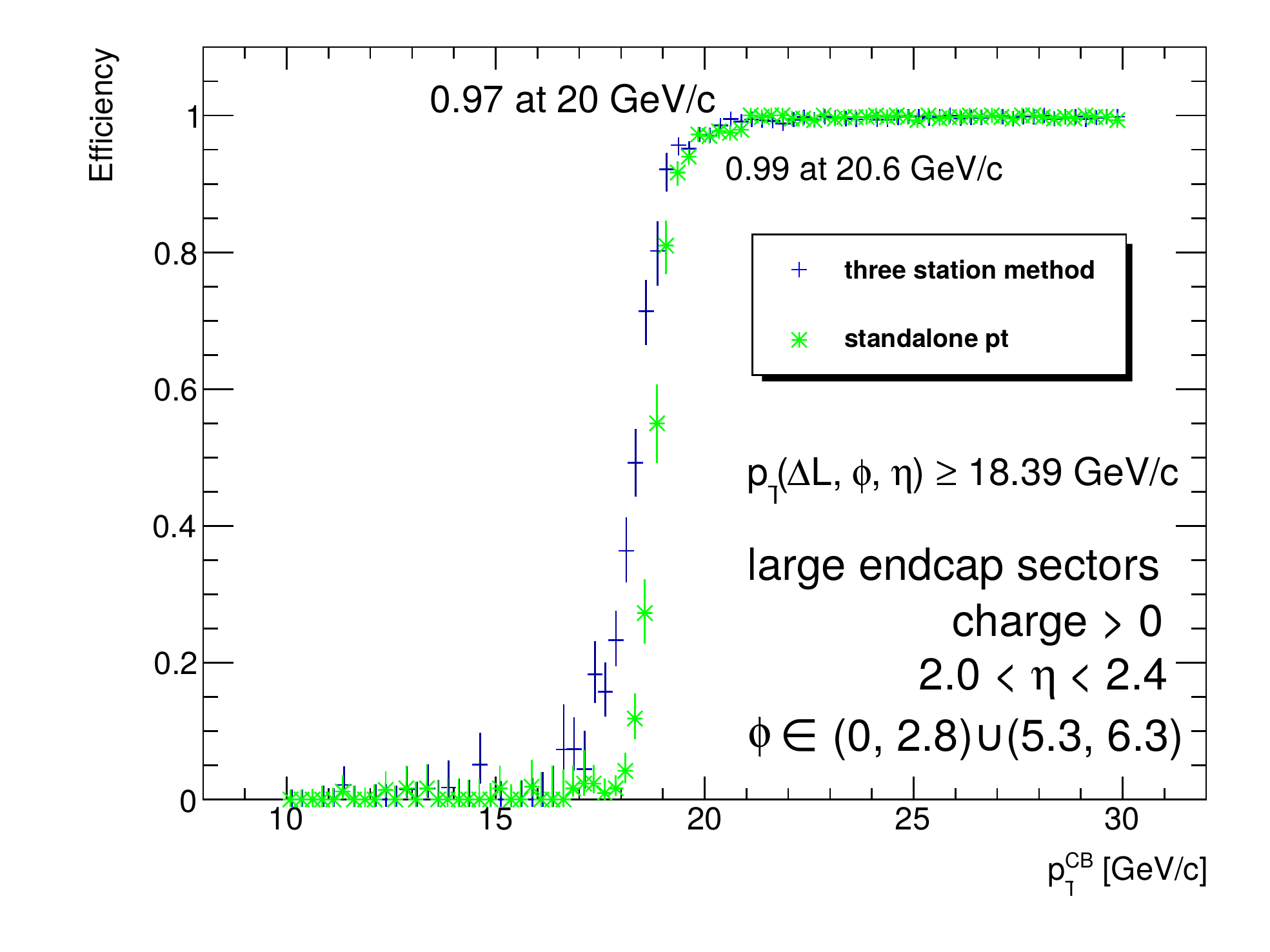}
\caption{Trigger efficiency for positively charged muons in large sectors at $2.0 < \eta < 2.4$ and $\phi \in (0,2.8)\cup(5.28,6.28)$ as function of the muon transverse momentum $p_\mathrm{T}^\mathrm{CB}$ reconstructed using the combined algorithm. A threshold of $p_\mathrm{T}^\mathrm{SA} = 18.39\,\mathrm{GeV}$ has been applied using the muon transverse momentum given by the parametrisation $p_{\mathrm{T}}(\Delta L,\phi,\eta)$ and by the energy-loss corrected standalone algorithm.}
\label{fig:5_efficiency_endcap}
\end{figure}



The resolution of the three station method is determined by taking the width of the distribution $(p_\mathrm{T}^{CB} - p_\mathrm{T}(\Delta s, \phi, \eta))/p_\mathrm{T}^{CB}$.
In Figures~\ref{fig:5_resolution_barrel} and \ref{fig:5_resolution_endcap} the distributions for the barrel region and the end-cap region are shown. The resolution is found to be $\sigma = 0.0410 \pm 0.0003$ and $\sigma = 0.0269 \pm 0.0009$ respectively.
The improved resolution in the end-cap region with respect to the barrel region can be partially attributed to correlations between the combined and stand-alone reconstruction algorithms. Another contribution is the lower energy loss of muons in large barrel sectors compared to end-cap sectors, resulting in a better momentum resolution in the end-cap region due to less pronounced non-Gaussian tails.

Setting a threshold of $p_\mathrm{T}^\mathrm{thresh} = 17.55\,\textrm{GeV}$ in the barrel region, one arrives at an efficiency curve shown in Figure~\ref{fig:5_efficiency_barrel}. For comparison, the curve obtained by using the standalone momentum is also shown. At \unit{20}{GeV} the  trigger efficiency is 0.95, while the full efficiency is reached at \unit{23.9}{GeV}.

The efficiency curve in the end-cap region is shown in Figure~\ref{fig:5_efficiency_endcap}. It is steeper and almost fully efficient at \unit{20}{GeV}. Full efficiency is reached at \unit{20.6}{GeV}.
The threshold is chosen to be  $p_\mathrm{T}^\mathrm{thresh} = 18.39\,\textrm{GeV}$. Again for comparison the curve obtained by using the standalone momentum is overlaid. 
Both curves are very similar in shape, their different slope follows as a consequence of the aforementioned correlation and reduced energy loss in the end-cap region.

\section{Summary}
It has been shown that a MDT trigger can improve the selectivity of the first-level muon trigger significantly.
Based on an iterative fitting method, a parametrisation of the muon's transverse momentum $p_{\mathrm{T}}(\Delta s,\phi,\eta)$ can be determined.
For a threshold of $p_\mathrm{T}^\mathrm{thresh} = 17.55\,\textrm{GeV}$ in the barrel region, the trigger is $95\%$ efficient at the efficiency point $p_T^{\mathrm{eff}} = 20\,\mathrm{GeV}$ and becomes fully efficient at \unit{23.9}{GeV}.
In the end-cap region, setting a threshold of $p_\mathrm{T}^\mathrm{thresh} = 18.39\,\textrm{GeV}$, the trigger is $97\%$ efficient at $p_T^{\mathrm{eff}} = 20\,\mathrm{GeV}$ and becomes fully efficient at \unit{20.6}{GeV}.
\ifCLASSOPTIONcaptionsoff
  \newpage
\fi

\end{document}